\documentclass[a4paper]{jpconf}
\usepackage{graphicx}
\begin{document}
%\documentclass[11pt,twocolumn,twoside,a4paper,amsmath,amssymb,aps,showkeys,showpacs]{revtex4}
%
%\usepackage{amsfonts}
%\usepackage{graphics} %   <------- for LatTeX + EPS
%\usepackage{epsfig}
%\usepackage{fancyheadings}
%
%
%\textheight=22cm
%\textwidth=17.2cm
%\columnsep=0.8cm
%
%\pagestyle{fancy}
%\parskip 0pt
%\parindent 24pt
%\voffset2cm
%\def\baselinestretch{.9}
%
%
%\begin{document}
%\thispagestyle{myheadings}
%%%%%%%%%%%%%%%%%%%%%%%%%% Title %%%%%%%%%%%%%%%%%%%%%%%%%%%%%%%%%%%%%%
%\rhead[]{}%<------
%\lhead[]{}%<------
%\chead[David Kettler on behalf of the STAR collaboration]{Quadrupole Spectrum}%<------short title

\title{Azimuth Quadrupole Systematics in Au-Au Collisions}

\author{David Kettler (STAR Collaboration)}
\address{University of Washington, Seattle, USA}
\ead{dkettler@u.washington.edu}

%\author{Author 2}
%\altaffiliation[Also at ]{JIPNR, Minsk, Belarus}
%\email{grossmann@physik.uni-marburg.de}
%
%\affiliation{%
%Fachbereich Physik, Philipps-Universit\"at,  Renthof 6,
% D-35032 Marburg,  GERMANY }%
%
%
%\author{Author 3}
%\email{d.lohse@utwente.nl}
%
%\affiliation{ Department of Applied Physics, University of
% Twente,\\ 7500 AE Enschede,  THE NETHERLANDS  }
%
%\received{ ?????????? }

\begin{abstract}
We have measured $p_t$-dependent two-particle number correlations on azimuth and pseudorapidity for eleven centralities of $\sqrt{s_{NN}} = 62$ and 200~GeV Au-Au collisions at STAR. 2D fits to these angular correlations isolate the azimuth quadrupole amplitude, denoted $2 v_2^2 \{ 2D \} ( p_t )$, from localized same-side correlations. Event-plane $v_2 ( p_t )$ measurements within the STAR TPC acceptance can be expressed as a sum of the azimuth quadrupole and the quadrupole component of the same-side peak. $v_2 \{ 2D \} ( p_t )$ can be transformed to reveal quadrupole $p_t$ spectra which are approximately described by a fixed transverse boost and universal L\'evy form nearly independent of centrality. A parametrization of $v_2 \{ 2D \} ( p_t )$ can be factored into centrality and $p_t$-dependent pieces with a simple $p_t$ dependence above 0.75 GeV/c. Results from STAR are compared to published data and model predictions.
\end{abstract}
%\pacs{ 25.75.Ag, 25.75.Gz, 25.75.Ld, 25.75.Nq }

%\keywords{elliptic flow, hydrodynamics, RHIC, quadrupole, minijets, correlations
%}

%\maketitle

%\renewcommand{\thefootnote}{\fnsymbol{footnote}}
%{\footnotetext[1]{Also delivered by SGn as evening (after dinner)
%talk at the ITI Conference in Bad Zwischenahn, October 2005}

%\renewcommand{\thefootnote}{\roman{footnote}}

%*****************   The Body of the Article:   *************************

\section{Introduction}
\label{introduction}

The collective behavior of particles produced in heavy-ion collisions is one of the major topics of study at RHIC, and a large quadrupole component in the distribution on azimuthal angle $\phi$ has been observed \cite{starreview}.  The large quadrupole azimuthal distribution in non-central collisions is usually described by elliptic flow, in which the different pressure gradients in and out of plane turn the initial position-space anisotropy of the colliding nuclei into a momentum-space anisotropy \cite{hydrointro}.

%***Do I even need this?  Not really the subject of this talk***
%This behavior has been described in terms of hydrodynamics via a pressure-driven expansion \cite{hydrointro}.  This description requires fast local thermalization of the hot, dense matter, initial conditions to describe the pre-equilibrium state, and final conditions to describe the freezeout into hadrons.

%The primary signal of this hydrodynamic expansion is radial flow: the collective transverse velocity of particles produced in the collision.

The azimuthal anisotropy is typically measured with the quantity $v_2$, the second Fourier component of the distribution of particle emission on azimuth with respect to the angle of the reaction plane \cite{basicv2}.  In real events this reaction plane angle is not known so many methods exist to estimate $v_2$.
%which will be explained more in Section~\ref{correlation}
The present analysis uses a method based on fitting two-dimensional angular correlations similar to \cite{quadrupole} but expanded to include the dependence on transverse momentum as well as centrality.  A simultaneous analysis of centrality and transverse momentum dependence of $v_2$ in the absence of
``nonflow'' from the same-side 2D peak (jet-like correlations) is presented.
%significant `nonflow'---correlations unrelated to the reaction plane---is presented.
%with significant implications for hydrodynamic descriptions of nuclear collisions.

%One of the earliest and still the most widespread methods
One method of measuring $v_2$ is the event-plane method, $v_2 \{ \textrm{EP}\}$, in which particles in the region of interest are used to estimate the reaction plane \cite{basicv2}.  A related approach, $v_2 \{ 2\}$, calculates $v_2$ directly from two-particle correlations 
%as the mean of $\cos ( 2 [ \phi_{i} - \phi_{j} ] )$
\cite{cumulants}.
%In \cite{azstruct} it was demonstrated that these two methods are nearly equivalent.
%---up to a simple approximation---different ways of expressing the same thing.
These two methods are essentially equivalent ($v_2 \{ \textrm{EP}\} \sim v_2 \{ 2 \}$ to 5\%) \cite{azstruct}.
This analysis, denoted by $v_2 \{ \textrm{2D} \}$, uses an approach similar to $v_2 \{ 2\}$ but includes the dependence of the two-particle distribution on pseudorapidity as well as azimuth.  Significant jet-like correlations are observed at all centralities \cite{mikeQM, fragevolve} and their 2D angular dependence can be used to distinguish them from the azimuth quadrupole component, which within the STAR TPC acceptance has no significant $\eta$ dependence \cite{v2EP}.  Jet-like correlations are separated from $v_2 \{ \textrm{2D} \}$ by 2D model fits described in Sec.~\ref{fitmodel}.

%Multiparticle methods, such as the four-particle cumulant and Lee-Yang zero methods are useful for reducing nonflow contributions but have difficulties in the more central collisions where $v_2$ values are small and in peripheral collisions where the number of particles is small \cite{cumulants, LYZ}.  It has also been suggested that large event-by-event fluctuations in $v_2$--even at a given centrality--could result in a significant difference between $v_2$ measured with two-particle and multiparticle methods \cite{v2fluct}.

%\cite{tpcs} acceptance has no significant pseudorapidity dependence.
%These correlations can be easily constructed for all centralities.

%The quadrupole results for the most central collisions are of particular interest because
%data from multiparticle methods (e.g., four-particle cumulants \cite{cumulants}, Lee-Yang Zeros \cite{LYZ}) are not available even for $p_t$-integrated $v_2$ in 0-5\% collisions \cite{v2EP, starLYZ}.

%do not work well in these cases \cite{LYZ, v2EP, starLYZ}.

\section{Correlations}
\label{correlations}

Angular correlations without a trigger particle are constructed by considering all pairs of particles in an event.  For each particle the azimuth angle $\phi$, pseudorapidity $\eta$ and transverse momentum $p_t$ are measured, defining a six dimensional two-particle space.  Integrating over $p_t$ and projecting the angular variables onto their difference axes $\eta_\Delta \equiv \eta_1 - \eta_2$ and $\phi_\Delta \equiv \phi_1 - \phi_2$ yields 2D angular autocorrelations \cite{quadrupole, mikeQM, mjdef}.
%The goal of this analysis is to incorporate the $p_t$-dependence, which will be discussed in more detail in Section~\ref{distributions}.

The correlation measure is constructed from pair densities.  Sibling pairs from the same event contain the primary signal.  Mixed pairs from different events are useful as a reference for removing detector artifacts.  The basic correlation measure we use is $\Delta \rho / \rho_{ref}$, the ratio of the sibling minus the mixed pairs to the mixed pairs, a per-pair measure.  It is also sometimes desirable to use a two-particle measure like $\Delta \rho / \sqrt{\rho_{ref}}$ \cite{fluctinv}.  However, this analysis uses $\Delta \rho / \rho_{ref}$ which is easily converted to $v_2$:
$2 v_{2}^{2} \{ 2D \} \equiv \Delta \rho [2] / \rho_{ref}$, 
%\begin{eqnarray}
%v_2 \{ 2D \} \equiv \sqrt{\frac{1}{2} \frac{\Delta \rho}{\rho_{ref}} [2]},
%\label{eqn:v2}
%\end{eqnarray}
where the $[2]$ indicates the quadrupole term of the fit model in Eq.~\ref{eqn:fitfunc} below.

%\section{Transverse Momentum Dependence}
%\label{distributions}

When we consider momentum dependence it is desirable to use transverse rapidity instead of transverse momentum:
%(or mass $m_t$), defined as
$y_t = \ln \left\{ (p_t + m_t ) / m_0 \right\}$.
Transverse rapidity is the analogue to longitudinal rapidity but in the transverse direction.
%Its desireable scaling properties means that we can make bins that are uniform in $y_t$.
%Since we are usually measuring unidentified particles
For unidentified particles we make the approximation $m_0 = m_{\pi}$
%because pions are the most numerous particle produced in nuclear collisions and for
%which for most of the $p_t$
%range of interest
%ranges we will be discussing the choice of mass
%will not have a significant effect on $y_t$.
%In this range $y_t$ is approximately $\ln ( p_t )$ but it is well-defined even for $p_t = 0$.
to define a regularized $\ln ( p_t )$ measure.

%The primary goal of this study is to understand the $y_t$ dependence of these correlations.
%The $y_t$-dependence of these correlations is studied by making cuts on $y_t$ and examining the histograms in a restricted interval.  In a two-particle correlations it is possible to restrict the transverse momentum of each particle independently so the $y_t$-dependence is inherently two-dimensional.  While the two-dimensional dependence is interesting in its own right, we will construct a one-dimensional function on $y_t$ to compare to published $v_2 ( p_t )$ data.

\begin{figure*}
\resizebox{1.00\textwidth}{!}{
  \resizebox{1.00\textwidth}{!}{\includegraphics{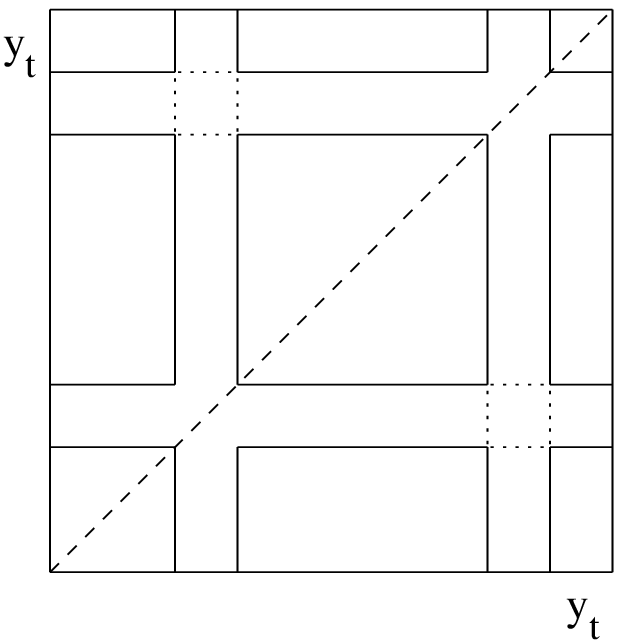}}
  \includegraphics{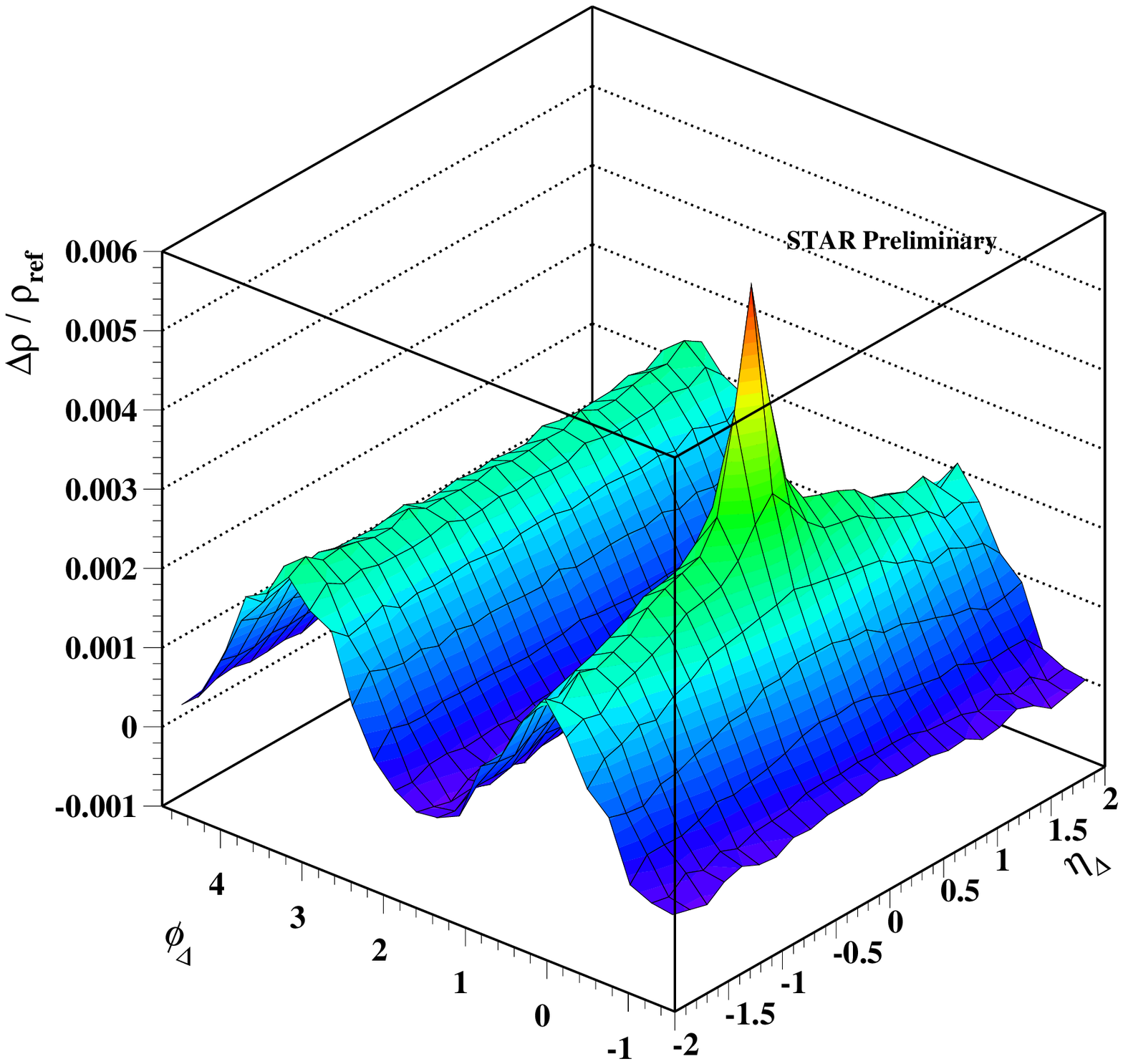}
  \includegraphics{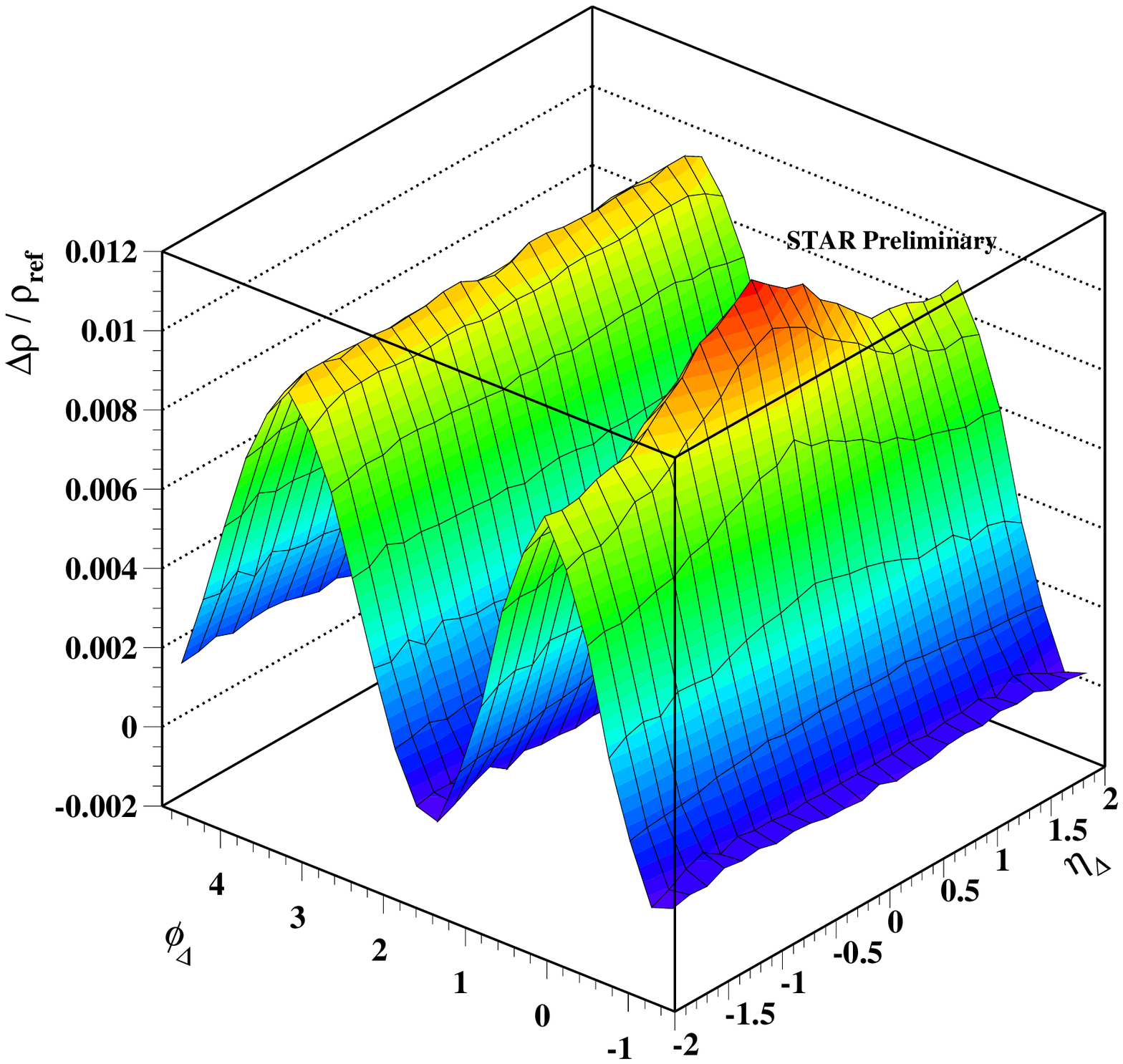}
  \includegraphics{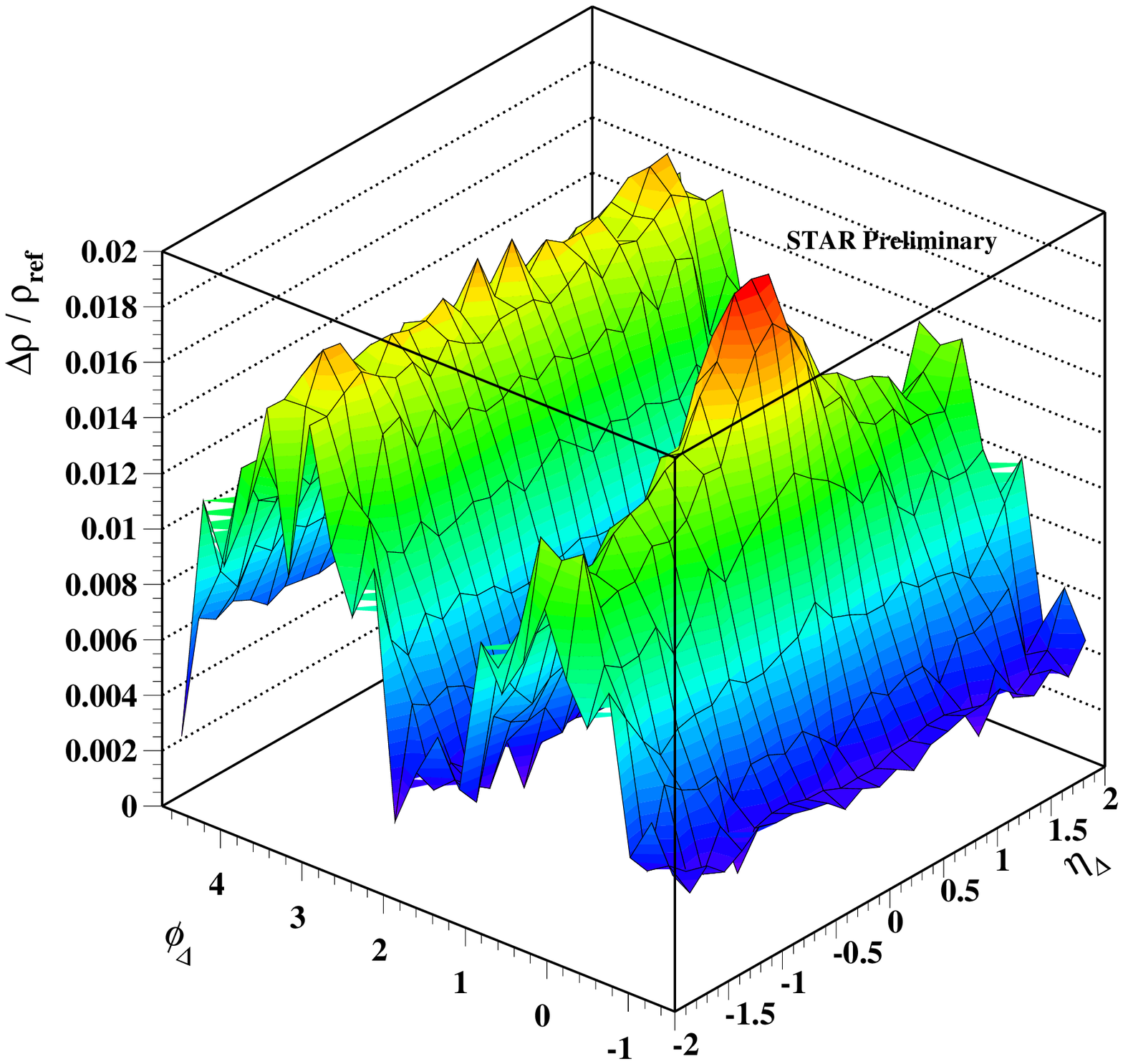}
}
\caption{\label{fig:examples}Left Panel: Two examples of marginal distribution cuts in $y_t \times y_t$ space.  Remaining panels: An example of the $y_t$ evolution of correlation structures for 40-50\% central collisions.  The plots correspond to $y_t$ bins of $1.4 < y_t < 1.8$, $3.0 < y_t < 3.4$, and $3.8 < y_t < 4.2$.}
\end{figure*}

We construct marginal distributions by restricting the $y_t$ range of one of the particles in the two-particle correlation to a certain interval.
% while the other is unrestricted.
Due to the inherent diagonal symmetry in $y_t \times y_t$ space this corresponds to making cross-shaped cuts as seen in Fig.~\ref{fig:examples} (left panel).  These cuts are analogous to the procedure in a standard event-plane $v_2 ( p_t )$ analysis \cite{basicv2} in which particles from the full momentum range are used to estimate the event plane and particles in the momentum interval are correlated to it.  The marginal distributions allow us to study correlations at large $p_t$ values
%than a joint distribution on $( y_{t1} , y_{t2} )$
 and are easily converted into the usual $v_2 ( p_t )$ measure for comparison with published $v_2 ( p_t )$ data.

%\begin{figure}
%\resizebox{0.50\textwidth}{!}{
%  \includegraphics{ytytcuts.eps}
%  \includegraphics{ytytcuts.eps}
%}
%\caption{\label{fig:marginal} Left Panel: Examples of marginal distribution cuts in $y_t \times y_t$ space.  Right Panel: Relationship of integrated $v_2$ to $v_2 (p_t )$.}
%\end{figure}

In this analysis there is a minimum $p_t$ cut for all particles of $150 \textrm{ MeV}/c$ which corresponds to a pion $y_t$ of 0.93.  There are nine $y_t$ bins.  The first $y_t$ bin includes all particles with $y_t$ less than 1.4 and the seven subsequent bins are 0.4 units of $y_t$ wide.  The final bin includes all particles with $y_t$ greater than 4.2.  This corresponds to particles above $\sim 5 \textrm{ GeV}/c$.

%We can express the $p_t$-integrated $v_2$ at centrality $b$ in terms of the $p_t$-dependent $v_2$ by 
%\begin{eqnarray}
%v_{2} (b) = \frac{1}{\rho_0 (b)} \int dp_t p_t \rho_0 (p_t , b) v_2 (p_t , b),
%\label{eqn:v2pt}
%\end{eqnarray}
%where $\rho_0 (p_t , b)$ is the single-particle spectrum \cite{tcspectra}.  As the spectrum falls off exponentially at larger $p_t$, integrated $v_2$ numbers are heavily weighted toward lower $p_t$ particles and
%%This is shown explicitly in Fig.~\ref{fig:marginal} (right panel).
%%As integrated results are more readily available for many $v_2$ methods and centralities than the proper $p_t$-dependent $v_2$ they are sometimes used to renormalize the functional form of differential results from more readily available methods--such as $v_2 \{ \textrm{EP} \}$--for background subtraction \cite{***something?***}.
%%But as can be seen in Fig.~\ref{fig:marginal}
%%The integrated results have virtually no sensitivity beyond the first few $y_t$ bins and should not be used to conclude anything about the behavior of $v_2$ at higher $p_t$.
%have virtually no sensitivity beyond the first few $y_t$ bins.  Integrated results should not be used to conclude anything about the behavior of $v_2$ at higher $p_t$.

\section{Fit Parameters}
\label{fitmodel}

This analysis is based on 14.5 million \mbox{Au-Au} collisions at $\sqrt{s_{NN}} = 200 \textrm{ GeV}$ and 6.7 million \mbox{Au-Au} collisions at $\sqrt{s_{NN}} = 62 \textrm{ GeV}$ observed with the STAR
%time projection chamber \cite{tpcs}
TPC.  The acceptance was defined by transverse momentum $p_t > 0.15 \textrm{ GeV}/c$, $| \eta | < 1$ and $2\pi$ azimuth.
%\mbox{Au-Au} collision centrality was defined as in \cite{centralities}.
Minimum-bias event samples were divided into 11 centrality bins: nine $\sim 10\%$ bins from near 100\% to 10\%, the last 10\% divided into two $\sim 5\%$ bins.

\begin{figure*}
%\resizebox{1.00\textwidth}{.15\textheight}{
\resizebox{1.00\textwidth}{!}{
  \includegraphics{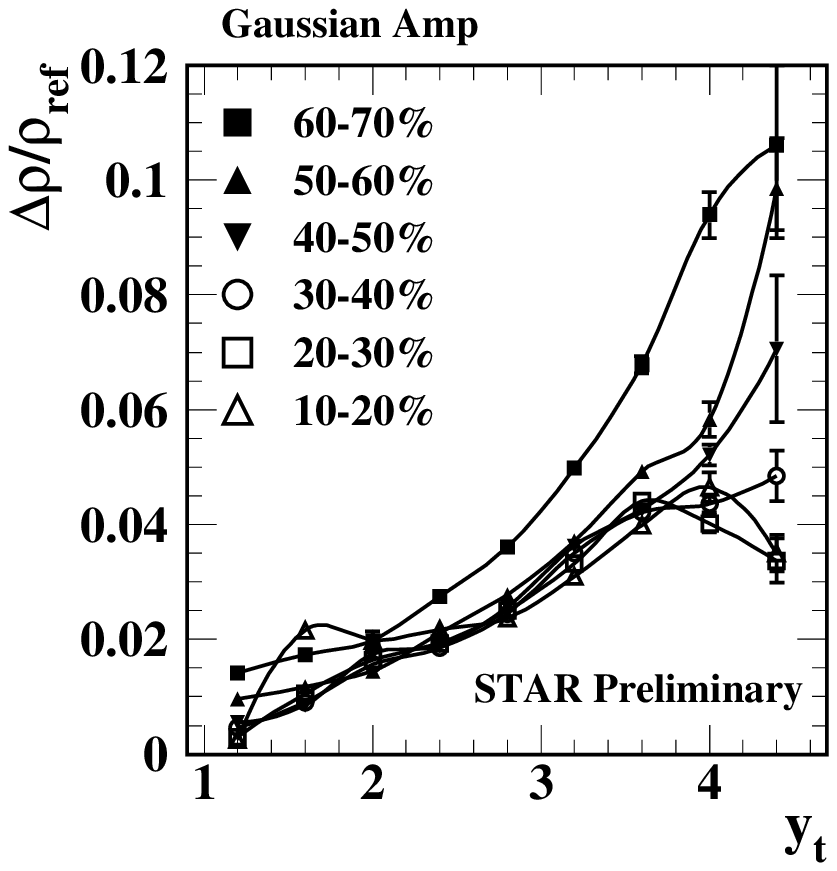}
  \includegraphics{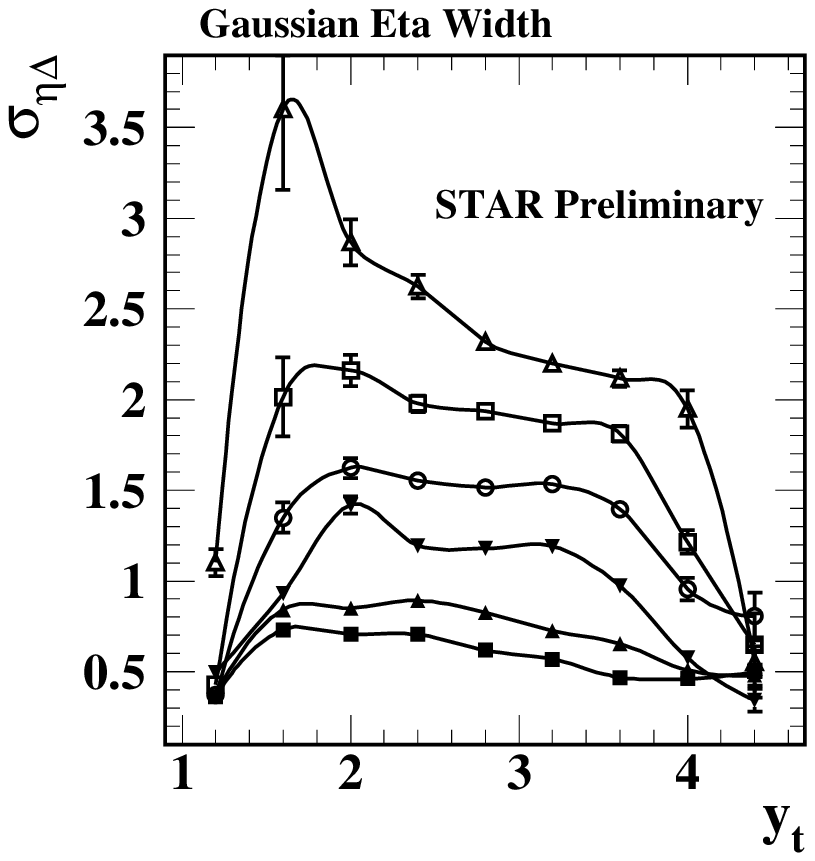}
  \includegraphics{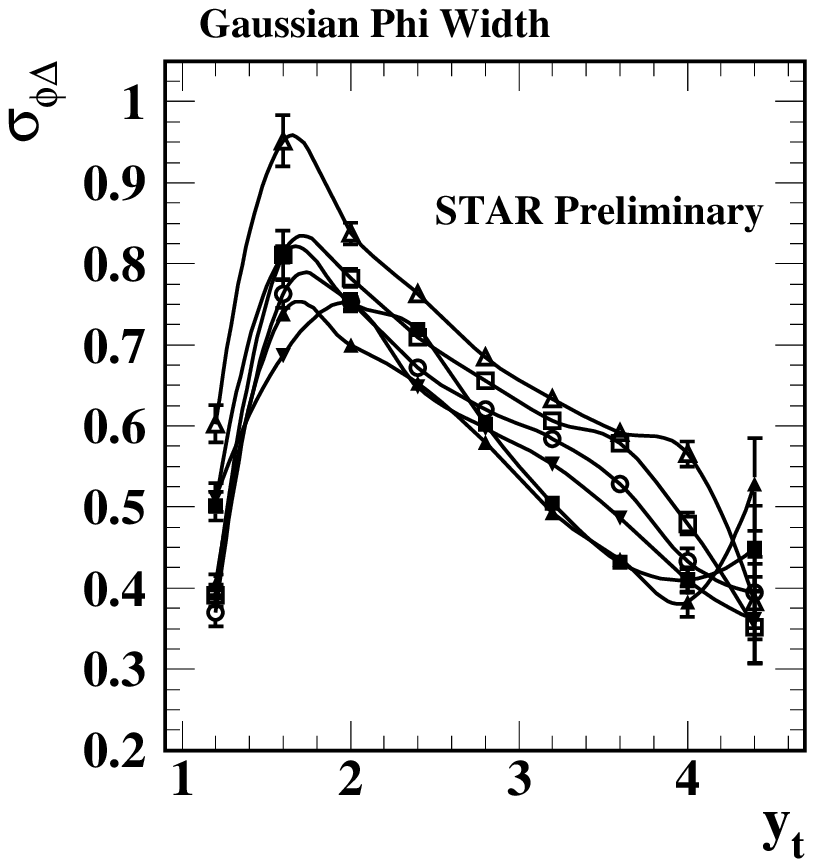}
  \includegraphics{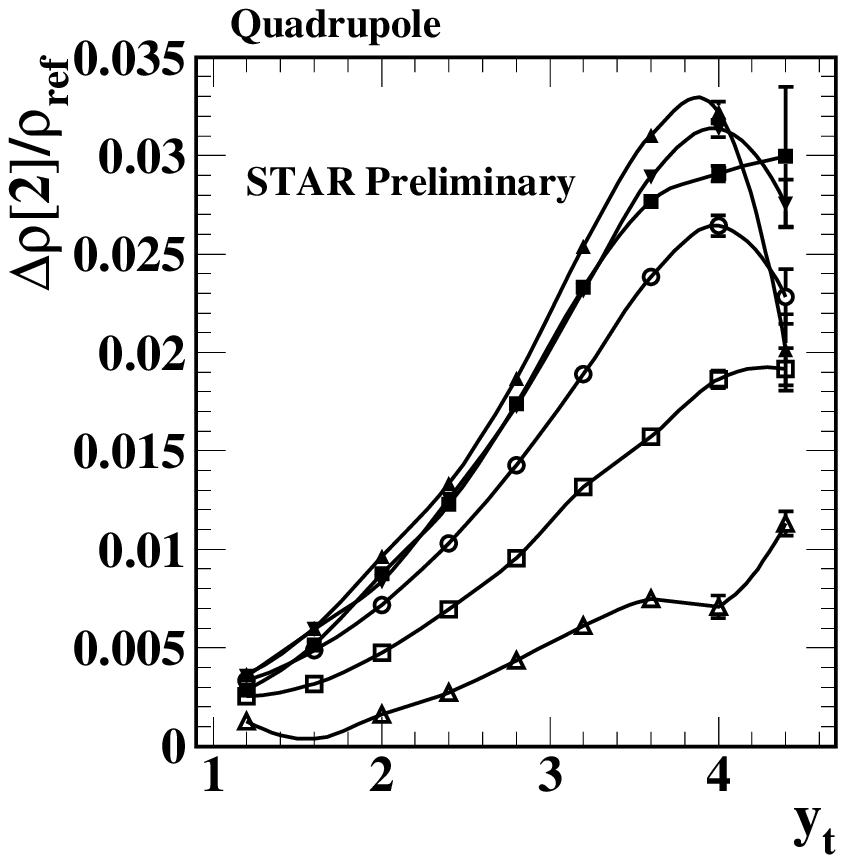}
}
\caption{\label{fig:fitparms}The amplitude, width in $\eta_\Delta$, and width in $\phi_\Delta$ of the 2D Gaussian and the amplitude of the quadrupole component of the fit for Au-Au 200~GeV 60-70\%, 50-60\%, 40-50\%, 30-40\%, 20-30\%, and 10-20\% collisions.  Error bars are for fit errors only.}
%and do not represent a detailed study of the systematics.}
\end{figure*}

Two-particle correlation histograms are constructed for each of the 11 centrality and 9 $y_t$ bins described above for a total of 99 bins at each $\sqrt{s_{NN}}$.  Example histograms
%for different $y_t$ values in 40-50\% central collisions
are shown in Fig.~\ref{fig:examples}.
%The fitting model is used to extract the relevant quantities from the histograms.
Fig.~\ref{fig:fitparms} shows the evolution of the most relevant fit parameters at $\sqrt{s_{NN}} = 200 \textrm{ GeV}$ with $y_t$ for six centrality bins.  Systematics errors are still under study.  62 GeV results have similar trends as 200 GeV.

In order to distinguish the quadrupole signal from $\eta$-localized (jet-like) correlations we will fit the major structures in the 2D histograms.  The fitting model used is similar to that used in \cite{quadrupole},
%In this case the exponential term that was used to model the sharp peak present in the angular correlations near the origin due to electron pair production has been removed and those bins are simply excluded from the fit.
but the sharp peak from electron pairs is excluded rather than fit.
The histograms are constructed with 24 bins on $\phi_{\Delta}$ with centers ranging from $-\pi / 2$ to $17 \pi / 12$ in steps of $\pi / 12$ and 25 bins on $\eta_{\Delta}$ with centers ranging from $-1.92$ to $1.92$.  A total of seven bins are excluded from the fit, where $\eta_\Delta = 0$ and $\phi_\Delta = 0, \pm \pi / 12$ and where $\phi_\Delta = 0$ and $\eta_\Delta = \pm 0.08 , \pm 0.16$.

The complete fitting function is then
\begin{eqnarray}
F \hspace{-.05in} &=& \hspace{-.05in} A_{\phi_\Delta} \cos (\phi_\Delta ) + A_{2 \phi_\Delta} \cos (2\phi_\Delta ) + A_0 e^{- \frac{1}{2} \left( \frac{\eta_\Delta}{\sigma_0} \right) ^2} 
+ A_1 e^{- \frac{1}{2} \left\{ \left( \frac{\phi_\Delta}{\sigma_{\phi_\Delta}} \right) ^2 + \left( \frac{\eta_\Delta}{\sigma_{\eta_\Delta}} \right) ^2 \right\} } + A_2 .
\label{eqn:fitfunc}
\end{eqnarray}
There are 5 terms and 8 parameters.  Two of the terms are sinusoids on $\phi_\Delta$, one is a constant offset, and one models only structure on $\eta_\Delta$.  Thus, the remaining term---the two-dimensional Gaussian---is solely responsible for describing ``nonflow'' \cite{azstruct}.
%This is made explicit by the comparison to published $v_2 \{ \textrm{EP}\}$ data shown in Fig.~\ref{fig:compare} (left and middle panels).
%For more central high-$p_t$ collisions the single 2D Gaussian is not a perfect model of the same-side structure, however the impact of the details of this structure on the quadrupole term is minimal.

%We also tried modeling the away-side structure with a 1D Gaussian rather than the negative dipole as the away-side peak should narrow at larger $p_t$.  However, at least within the $p_t$ ranges we explored there was no discernible difference in the fit quality when using a 1D Gaussian instead of a dipole so only the dipole results are presented here.

%The typical $\chi^2 / \textrm{n.d.f}$ values for the fits are $\sim 0.8$ to $2.5$ with the largest values occuring at high-$p_t$ for mid-central collisions.  The structures in the residuals indicate that the detailed shape of the same-side peak is not being properly modeled for all $p_t$.  This is consistent with triggered-particle analyses which indicate that at large-$p_t$ there are seperate 'jet' and 'ridge' components of the peak \cite{jetridge}.  However, these components are still subtle and difficult to distinguish in these marginal distributions.

\begin{figure*}
\resizebox{1.00\textwidth}{!}{
  \includegraphics{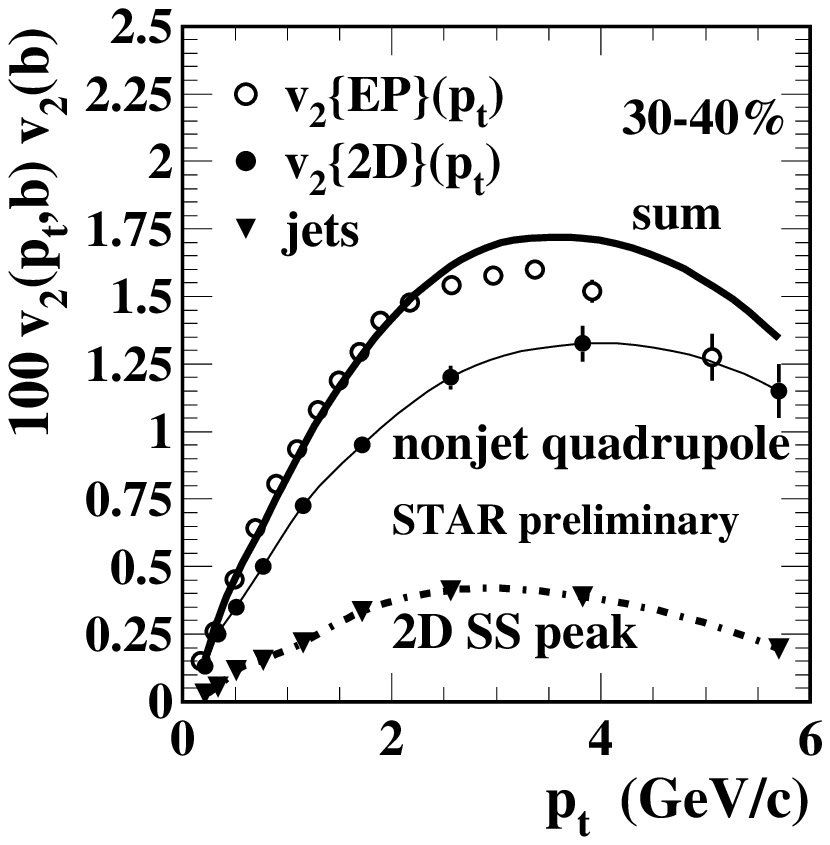}
  \includegraphics{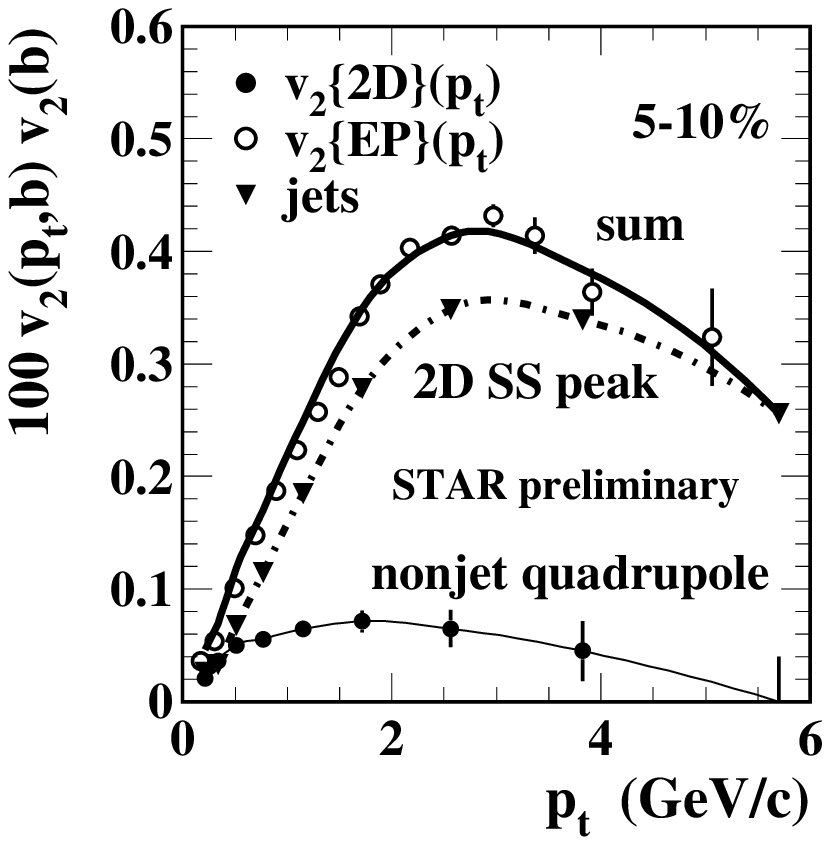}
  \includegraphics{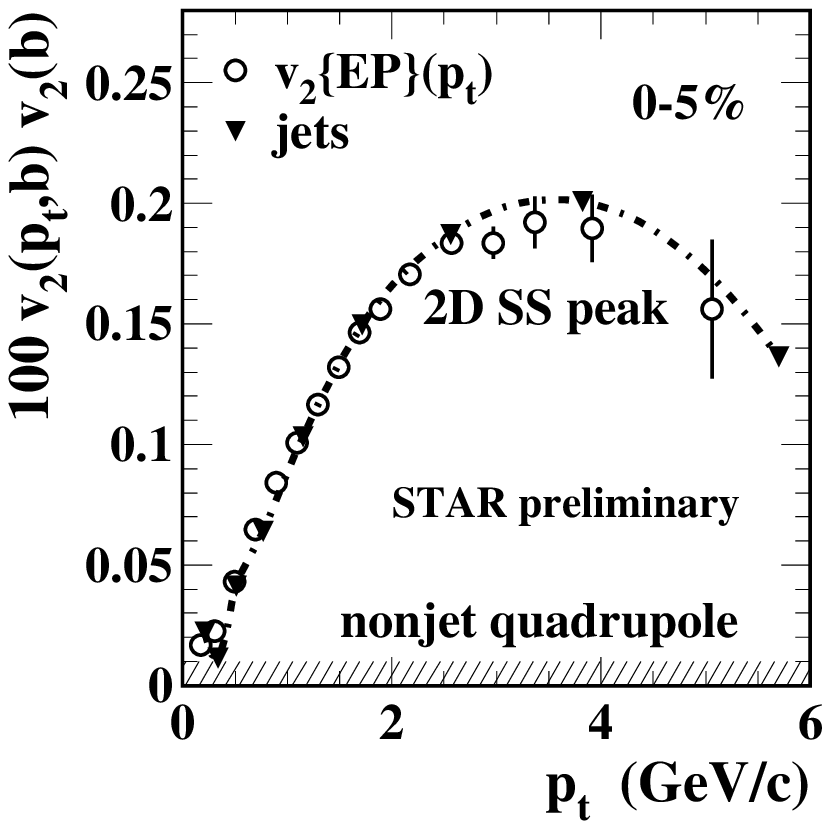}
  \includegraphics{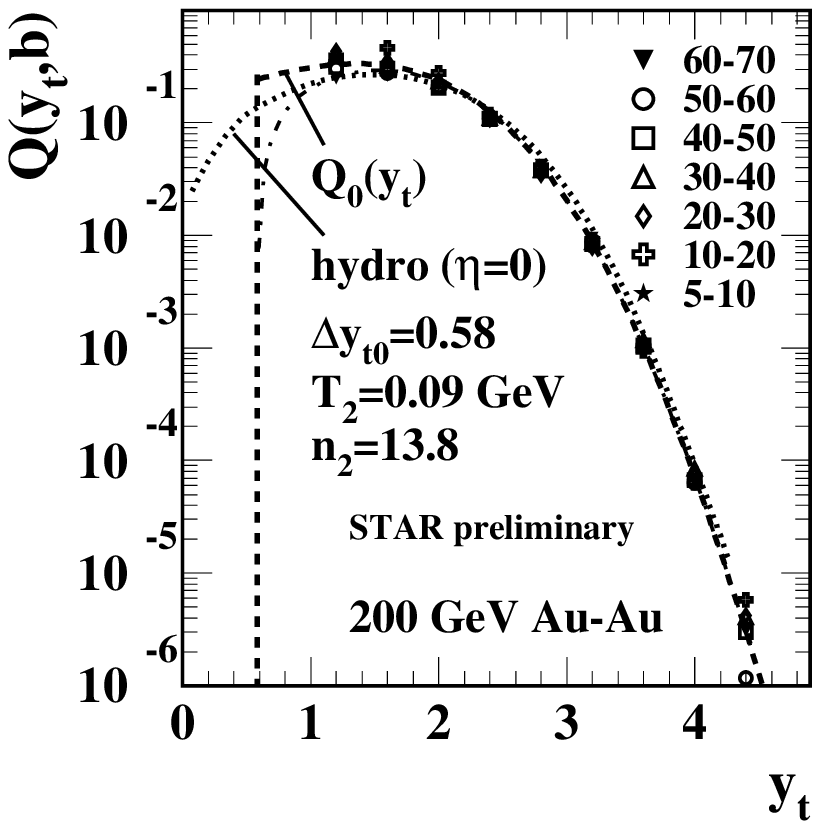}
}
\caption{\label{fig:compare} First three panels: Comparison of published $v_2 \{ \textrm{EP} \}$ for Au-Au 200 GeV 30-40\%, 5-10\% and 0-5\% central collisions \cite{v2EP} and results from this analysis.  Errors are fit errors only.  Right Panel: The quadrupole spectrum data $Q(y_t , b)$ and fit $Q_0 (y_t )$.  The dotted curve is a zero-viscosity limit of a viscous hydro theory defined in \cite{hydrotheory}.}
\end{figure*}

Fig.~\ref{fig:compare} (first three panels) shows measured quadrupole results from this analysis vs. $p_t$ compared to previously published $v_2 \{ \textrm{EP}\}$ results for 30-40\%, 5-10\% and 0-5\% central collisions.  In the case of 0-5\% the quadrupole component was found to be too small to measure reliably so the cross-hatched region represents an upper bound.
We calculate the contribution of the 2D Gaussian peak to the second Fourier component using the measured amplitudes and widths in the curve labeled 'jets'.
%The dotted curve represents a simplification that does not take into account the $p_t$-evolution of the widths of the 2D Gaussian peak.
We find
%that in the more central collisions
$v_2 \{ EP \}$ seems to be dominated by the SS peak structure in 5-10\% collisions and entirely due to it in 0-5\% collisions.  This has important implications for the interpretation of published $v_2$ data.

\section{Quadrupole Descriptions}
\label{qspec}

Consider a model in which the single-particle density on $y_t$ and $\phi$ and corresponding transverse boosts can be decomposed into azimuth-independent $\rho_0$ and quadrupole $\rho_2$ pieces \cite{quadspec}.  Define $V_2 (y_t , b) \equiv \rho_0 (y_t , b) v_2 (y_t , b)$ and invoke a blastwave model to get the relation
$V_2 (y_t , b) \approx p_t \Delta y_{t2} (b) \rho_2 (y_t , b) / (2 T_2)$
%\begin{eqnarray}
%V_2 (y_t , b) \approx \frac{p_t \Delta y_{t2} (b)}{2 T_2} \rho_2 (y_t , b) ,
%\label{eqn:blastwave}
%\end{eqnarray}
where $\Delta y_{t2}$ is the quadrupole component of the radial boost distribution, $\rho_2$ is the quadrupole spectrum, and $T_2$ is a blastwave parameter.
Now construct the unit-integral quantity $Q ( y_t , b ) \equiv [ V_2 (y_t , b) / p_t ] / [ V_2 ( b ) \langle 1 / p_t \rangle ] \approx [ \rho_2 (y_t , b) ] / [ \rho_2 (b) ] $.
%\begin{eqnarray}
%Q(y_t , b) \equiv \frac{V_2 (y_t , b)/p_t }{V_2 (b) \left< 1/p_t \right>} \approx \frac{\rho_2 (y_t , b)}{\rho_2 (b)} .
%\label{eqn:Qdef}
%\end{eqnarray}
Most of the parameters drop out and this variable directly relates measurable quantities to the quadrupole spectrum model.

%When $\rho_2 (y_t , b)$ is plotted in this form there
%does not appear to be any significant
%is little centrality dependence to the quadrupole spectrum.
There is little centrality dependence to the quantity $Q ( y_t , b )$.
A single L\'evy distribution $Q_0 (y_t)$ can describe the data within 10\% with a few simple parameters: a $y_t$ boost $\Delta y_{t0} =0.58$ and parameters $T_2 = 0.09 \textrm{ GeV}$ and $n_2 = 13.8$.  There is deviation from this curve for very central collisions at low-$p_t$.
%Detailed systematic errors are still under study.
A similar analysis was made in \cite{quadspec} for minimum-bias identified particles without centrality dependence and the observed boost values agree.
%The $\Delta y_{t0}$ values for the two cases seem to agree quite well.
It is notable that the boost is a single sharp value and not a distribution as would be expected from transverse Hubble expansion.
%Parameter $T_2$ differs significantly from the $T_0 \approx 0.145 \textrm{ GeV}$ that is inferred from the full single-particle spectrum \cite{tcspectra}.
%A detailed study of the systematics is in progress.

A related and more accurate parametrization has been made.  At higher $p_t$ it is observed that $v_2 \{\textrm{2D} \} ( p_t , b )$ is proportional to $p_t \exp ( - p_t / 4 )$ so we derive the form:
\begin{eqnarray}
\label{eqn:param}
v_2 \{\textrm{2D} \} ( p_t , b ) \approx \langle 1 / p_t \rangle v_2 \{\textrm{2D} \} ( b ) p_t \exp ( - p_t / 4 ) \times f ( p_t , b ) ,
\end{eqnarray}
where $\langle 1 / p_t \rangle \approx 2.1 \: ( \textrm{GeV}/c )^{-1}$ and $f ( p_t , b )$ is a dimensionless factor necessary for the full $p_t$ and centrality dependence.  It is fit to data to get
$f ( p_t , b) = 1 + C ( b ) [ \textrm{erf} ( y_t - 1.2 ) - \textrm{erf} ( 1.8 - 1.2 ) ]$
%\begin{eqnarray}
%\label{eqn:fcorrection}
%f ( p_t , b) = 1 + C ( b ) [ \textrm{erf} ( y_t - 1.2 ) - \textrm{erf} ( 1.8 - 1.2 ) ] ,
%\end{eqnarray}
and $C ( b ) = 0.12 - ( \nu - 3.4 ) / 5 - [ ( \nu - 3.4 ) / 2 ]^5$.

%\section{Conclusions}
%\label{conclusions}

Using 2D correlations we observe $v_2 ( p_t )$ for a wide range of centralities without significant ``nonflow'' contamination and conclude that published
$v_2 (p_t , b) \{ \textrm{EP} \}$ results \cite{v2EP} contain a significant contribution from the same-side peak correlations.

We have measured the quadrupole on $y_t$ as a function of $b$ and deduced the quadrupole spectrum variation with centrality.
This is the first complete factorization of the scaling properties of the quadrupole.
A detailed study of the systematics is in progress.
When the quadrupole spectrum is isolated we also observe a single $b$-independent $y_t$ boost value that is consistent with the minimum-bias results for different particle species \cite{quadspec}.
We developed an accurate parametriazation of $v_2 \{ \textrm{2D} \}$ over a large kinetic domain and surprisingly simple trends are observed.

\label{last}
\end{document}